\documentclass[12pt]{article}
\title{Expressions for two generalized Furdui series} 
\author{Mark W. Coffey\\
Department of Physics\\
Colorado School of Mines\\
Golden, CO  80401\\
(Received $\mbox{~~~~~~~~~~~~~~~~~~~~~~~~~~~~~~~2009}$)}
\date{October 7, 2009}  
\begin{document}
\maketitle
\baselineskip=25 pt
\begin{abstract}

We solve two problems of analysis and special function theory recently 
posed by Furdui.  The series in question are special cases in our solution.

\end{abstract}
 
\vspace{.25cm}
\baselineskip=15pt
\centerline{\bf Key words and phrases}
\medskip 

\noindent

Stieltjes constants, Gamma function, digamma function, polygamma function,
Riemann zeta function, Hurwitz zeta function, polylogarithm function

\vfill
\centerline{\bf 2000 AMS codes}
33B15, 11M06

\baselineskip=25pt
\pagebreak
\medskip
\centerline{\bf Statement of results}
\medskip

We let $\Gamma$, $\psi$, and $\psi^{(j)}$ denote the Gamma, digamma, and polygamma
functions, respectively \cite{andrews}.  We let $\gamma=-\psi(1)$ be the Euler constant.  We let $\zeta(z)$ denote the Riemann zeta function, $\zeta(z,a)$ the
Hurwitz zeta function, and Li$_s$ the polylogarithm function \cite{sri}.  The 
latter functions may be initially defined by the series 
$$\mbox{Li}_s(z)=\sum_{k=1}^\infty {z^k \over k^s}, ~~~~|z| \leq 1, \eqno(1)$$
and analytically continued through out the complex plane.  In the case of 
integral index, as occurs in the following, we also have an expression in terms
of the generalized hypergeometric function $_pF_q$ \cite{andrews}:
$$\mbox{Li}_n(z)=z ~_{n+1}F_n(1,1,\ldots,1;2,\ldots,2;z).  \eqno(2)$$
We have the special case
$$\mbox{Li}_1(z)=-\ln(1-z).  \eqno(3)$$

We then have
{\newline \bf Proposition 1}.  Put for integers $j\geq 0$ and $|z|\geq 1$, $z \neq -1$,
$$S_j(z)  \equiv \sum_{n=1}^\infty {{(-1)^n} \over z^n} {1 \over n^j}\left[
\zeta\left(1+{1 \over n}\right)-n-\gamma\right].  \eqno(4)$$
Then (a)
$$S_j(z)=\sum_{k=1}^\infty {{(-1)^k} \over {k!}} \gamma_k \mbox{Li}_{j+k}\left(
-{1 \over z}\right), \eqno(5)$$
and (b) (Furdui case \cite{furdui})
$$S_0(1) =\gamma_1 \ln 2+ \sum_{k=2}^\infty {{(-1)^k} \over {k!}} \gamma_k
(2^{1-k}-1)\zeta(k), \eqno(6)$$
where $\{\gamma_k\}_{k=0}^\infty$ are the Stieltjes constants for the Riemann zeta function \cite{coffey06,coffey09,stieltjes}.  

{\bf Proposition 2}.  Put for integers $j\geq 0$ and $|z|\geq 1$, $z \neq -1$,
$$T_j(z) \equiv \sum_{n=1}^\infty {{(-1)^n} \over z^n} {1 \over n^j}\left[n-
\Gamma\left({1 \over n}\right)-\gamma \right].  \eqno(7)$$
Let
$$\Gamma(x)-{1 \over x}=\sum_{j=0}^\infty {c_j \over {(j+1)!}}x^j, ~~~~~|x|<1, 
\eqno(8)$$
where $c_0=-\gamma$ and $c_1=\gamma^2+\zeta(2)$.  Then (a)
$$T_j(z)=-\sum_{k=1}^\infty {c_k \over {(k+1)!}} \mbox{Li}_{j+k}\left(
-{1 \over z}\right), \eqno(9)$$
and (b) (Furdui case \cite{furdui})
$$T_0(1)=-\sum_{k=2}^\infty {c_k \over {(k+1)!}} (2^{1-k}-1)\zeta(k)-{c_1 \over 2}
\ln 2.  \eqno(10)$$

{\bf Proposition 3}.  Let $\{\gamma_k(a)\}_{k=0}^\infty$ be the Stieltjes 
coefficients for the Hurwitz zeta function \cite{coffey06,coffey09,stieltjes}.
Put for integers $j\geq 0$, $\ell \geq 1$, $|z|\geq 1$, $z \neq -1$, and Re $a>0$,
$$S_{j\ell}(z,a)  \equiv \sum_{n=1}^\infty {{(-1)^n} \over z^n} {1 \over n^j}\left[
\zeta^{(\ell)}\left(1+{1 \over n},a\right)-(-1)^\ell n^{\ell+1}-(-1)^\ell \gamma_\ell (a)\right].  \eqno(11)$$
Then (a)
$$S_{j\ell}(z,a)=\sum_{k=\ell+1}^\infty {{(-1)^k} \over {(k-\ell)!}} \gamma_k(a) \mbox{Li}_{j+k-\ell}\left(-{1 \over z}\right), \eqno(12)$$
(b) for $j \geq 1$
$$S_{j\ell}(1,a)=\sum_{k=\ell+1}^\infty {{(-1)^k} \over {(k-\ell)!}} \gamma_k(a) (2^{1+\ell-j-k}-1)\zeta(j+k-\ell), \eqno(13)$$
and (c)
$$S_{0\ell}(1,a) =(-1)^\ell \gamma_{\ell+1} \ln 2+ \sum_{k=\ell+2}^\infty {{(-1)^k} \over {(k-\ell)!}} \gamma_k(a) (2^{1+\ell-k}-1)\zeta(k-\ell). \eqno(14)$$

{\bf Proposition 4}.  Put for integers $j\geq 0$, $\ell \geq 0$, and $|z|\geq 1$, 
$z \neq -1$,
$$U_{j\ell}(z)\equiv \sum_{n=1}^\infty {{(-1)^n} \over z^n}{1 \over n^j}\left[
\left({{\zeta'} \over \zeta}\right)^{(\ell)}\left(1+{1 \over n}\right)+(-1)^\ell
\ell! n^{\ell+1}+\ell! \eta_\ell\right].  \eqno(15)$$
Let
$${{\zeta'} \over \zeta}(s)=-{1 \over {s-1}}-\sum_{j=0}^\infty \eta_j(s-1)^j,
~~~~~~|s-1|<3,  \eqno(16)$$
where $\eta_0=-\gamma$ and $\eta_1=\gamma^2+2\gamma_1$ \cite{coffey09cam},
\cite{coffey04} (Appendix).
Then we have (a)
$$U_{j\ell}(z)=-\sum_{k=\ell+1}^\infty {{k!} \over {(k-\ell)!}} \eta_k \mbox{Li}_
{j+k-\ell}\left(-{1 \over z}\right), \eqno(17)$$
(b) for $j \geq 1$
$$U_{j\ell}(z)=-\sum_{k=\ell+1}^\infty {{k!} \over {(k-\ell)!}} \eta_k(2^{1+\ell-j
-k}-1)\zeta(j+k-\ell), \eqno(18)$$
and (c)
$$U_{0\ell}(1)=\eta_{\ell+1}\ln 2-\sum_{k=\ell+2}^\infty {{k!} \over {(k-\ell)!}} \eta_k(2^{1+\ell-j-k}-1)\zeta(j+k-\ell). \eqno(19)$$

\medskip
\centerline{\bf Proof of Propositions}
\medskip

{\it Proposition 1}.  We make use of the well known Laurent expansion
\cite{coffey06,coffey09,stieltjes}
$$\zeta(s)={1 \over {s-1}}+\sum_{k=0}^\infty {{(-1)^k \gamma_k} \over
k!} (s-1)^k, ~~~~~~ s \neq 1, \eqno(20)$$
where $\gamma_0=\gamma$.  Then we have
$$S_j(z)=\sum_{n=1}^\infty {{(-1)^n} \over z^n} {1 \over n^j}\sum_{k=1}^\infty
{{(-1)^k} \over {k!}} {\gamma_k \over n^k}$$
$$=\sum_{k=1}^\infty {{(-1)^k} \over {k!}} \gamma_k \sum_{n=1}^\infty {{(-1)^n}
\over {z^n n^{j+k}}}$$
$$=\sum_{k=1}^\infty {{(-1)^k} \over {k!}} \gamma_k \mbox{Li}_{j+k}\left(
-{1 \over z}\right), \eqno(21)$$
wherein we used the series definition (1).  For part (b) we use the alternating 
zeta function case
$$\mbox{Li}_k(-1)=(2^{1-k}-1)\zeta(k), \eqno(22)$$
together with the easily verified limit
$$\lim_{x\to 1} (2^{1-x}-1)\zeta(x)=-\ln 2.  \eqno(23)$$
Alternatively, we could make use of the special case (3) in Eq. (21).

{\bf Remarks}.  Numerically, we have $S_0(1)\simeq -0.0462635927840$ and
$\gamma_1 \ln 2 \simeq$ $-0.0504720979971$.  

As many series and integral representations for $\gamma_k$ are known, (e.g.
\cite{coffey06,coffey09}) (5) and (6) may be rewritten in a variety of ways.

By the functional equation of the zeta function, the summand of (4) could be
written in terms of $\zeta(-1/n)$.

{\it Proposition 2}.  This Proposition follows similarly, using the defining
expansion (8) for the constants $c_j$.  For part (b), we again use the case
(22) and the limit (23).

{\bf Remarks}.  Numerically, $T_0(1) \simeq 0.371990830350$ and 
$-c_1 (\ln 2)/2 \simeq -0.685561374577$.   

As a first approximation, one may take $c_k/(k+1)! \simeq (-1)^{k+1}$ for all
$k \geq 2$.

The constants $c_j$ may be systematically found from polygammic constants in
terms of Bell polynomials.  This is because $\Gamma'=\Gamma \psi$ and we may
appeal to Lemma 1 of \cite{coffeyutil}.

{\it Proposition 3}.  We have from \cite{coffey06,coffey09,stieltjes}
$$\zeta(s)={1 \over {s-1}}+\sum_{k=0}^\infty {{(-1)^k} \over k!} \gamma_k(a)(s-1)^k, ~~~~~~ s \neq 1, \eqno(24)$$
where $\gamma_0(a)=-\psi(a)$, for $\ell \geq 1$
$$\zeta^{(\ell)}(s,a)={{(-1)^\ell \ell!} \over {(s-1)^{\ell+1}}} +\sum_{k=\ell}^\infty {{(-1)^k} \over k!}  \gamma_k(a) k(k-1)\cdots (k-\ell+1)(s-1)^{k-\ell}, ~~~~~~
s \neq 1, \eqno(25)$$
Therefore, we have
$$\zeta^{(\ell)}\left(1+{1 \over n},a \right)-(-1)^\ell \ell! n^{\ell+1}-(-1)^\ell
\gamma_\ell(a) = \sum_{k=\ell+1}^\infty {{(-1)^k} \over {(k-\ell)!}}{{\gamma_k(a)} \over n^{k-\ell}}, \eqno(26)$$
giving
$$S_{j\ell}(z,a)=\sum_{k=\ell+1}^\infty {{(-1)^k} \over {(k-\ell)!}}\gamma_k(a)
\sum_{n=1}^\infty {{(-1)^n} \over z^n} {1 \over n^{j+k-\ell}}$$
$$=\sum_{k=\ell+1}^\infty {{(-1)^k} \over {(k-\ell)!}} \gamma_k(a) \mbox{Li}_{j+k-\ell}\left(-{1 \over z}\right). \eqno(27)$$
This proves part (a).  For part (b) we use relation (22).  For part (c) in
turn we use the limit (23).

{\it Proposition 4}.  We have from (16) 
$$\left({{\zeta'} \over \zeta}\right)^{(\ell)}(s)=-{{(-1)^\ell \ell!} \over
{(s-1)^{\ell+1}}}-\sum_{j=\ell}^\infty \eta_j j(j-1)\cdots(j-\ell+1)(s-1)^{j-\ell},
~~~~~~|s-1|<3,  \eqno(28)$$
giving
$$\left({{\zeta'} \over \zeta}\right)^{(\ell)}\left(1+{1 \over n}\right)
+(-1)^\ell \ell! n^{\ell+1}+\ell! \eta_\ell=-\sum_{j=\ell+1}^\infty {{j!} \over 
{(j-\ell)!}} {\eta_j \over n^{j-\ell}}.  \eqno(29)$$
Then we find
$$U_{j\ell}(z)=-\sum_{k=\ell+1}^\infty {{k!} \over {(k-\ell)!}}\eta_k \sum_{n=1}^\infty {{(-1)^n} \over z^n} {1 \over n^{k+j-\ell}}$$
$$=-\sum_{k=\ell+1}^\infty {{k!} \over {(k-\ell)!}} \eta_k \mbox{Li}_
{j+k-\ell}\left(-{1 \over z}\right). \eqno(30)$$
For part (b) we may use (22) and for part (c) (23).

{\bf Remarks}.  A known recursion relation \cite{coffey04} (Appendix) 
systematically gives the $\eta_j$ constants in terms of the Stieltjes constants.

Numerically we have $\eta_1 \ln 2 \simeq -0.129997$ and $U_{00}(1)\simeq 0.0975567$.

Similarly we may generalize Proposition 2 to sums containing derivatives of the $\Gamma$ function,
$$T_{j\ell}(z) \equiv -\sum_{n=1}^\infty {{(-1)^n} \over z^n} {1 \over n^j}\left[
-(-1)^\ell \ell! n^{\ell+1}+\Gamma^{(\ell)}\left({1 \over n}\right)-{c_\ell \over {\ell+1}}\right]$$
$$=-\sum_{k=\ell+1}^\infty {c_k \over {(k+1)}}{1 \over {(k-\ell)!}} \mbox{Li}_{j+k-\ell}\left(-{1 \over z}\right). \eqno(31)$$

Moreover, we may extend our method to sums with other analytic function summands, including for instance $\zeta^2+\zeta'-2\gamma \zeta$ and $\zeta^2-(\zeta'/\zeta)'-2\gamma \zeta$.  We could also similarly perform sums
over derivatives of the Lerch zeta function $\Phi$.

\pagebreak

\end{document}